\documentclass{birkmult}
\usepackage[utf8]{inputenc}
\usepackage[T1]{fontenc}
\usepackage[russian,english]{babel}
\usepackage{color} 
\usepackage{xcolor} 
\usepackage{amssymb}
\usepackage{amsmath}
\usepackage{dsfont} 
\usepackage{bbold} 
\usepackage{relsize} 
\usepackage[cmtip,arrow]{xy}
\usepackage{pb-diagram,pb-xy}
\usepackage{hyperref}
\usepackage{float}
\usepackage{cite}

\def\cmath{\color{blue}}

\newcommand{\Ali}[1]{{\cmath\begin{align*}#1\end{align*}}}

\def\bmat{\begin{pmatrix}}
\def\emat{\end{pmatrix}}
\newcommand{\barket}[1]{\left|#1\right\rangle} 
\newcommand{\brabar}[1]{\left\langle#1\right|} 
\newcommand{\braket}[1]{\left\langle#1\right\rangle} 
\def\bitand{\&} 
\def\bitnot{\sim} 
\def\N{\mathbb{N}} 
\def\C{\mathbb{C}} 
\def\Q{\mathbb{Q}} 

\newcommand{\cabs}[1]{\left|#1\right|} 
\def\denmat{\mathlarger{\rho}}

\ifx\e\undefined\DeclareMathOperator{\e}{e}\fi
\DeclareMathOperator{\Exp}{Exp} 
\newcommand{\farg}[1]{\!\left(#1\right)} 
\def\Hspace{\mathcal{H}} 
\DeclareMathOperator{\idmat}{\mathds{1}}
\def\imu{\mathrm{\mathbf{i}}} 
\newcommand{\inner}[2]{\left\langle#1\mid#2\right\rangle} 
\newcommand{\innerform}[3]{\braket{#1\cabs{#2}#3}} 
\DeclareMathOperator*{\KroneckerProduct}{\otimes}
\def\lcycle{\ell}	
\def\markstd{\!\star}
\newcommand{\Math}[1]{$\cmath{}#1$} 
\newcommand{\MathEq}[1]{\begin{equation*}\cmath{#1}\end{equation*}}
\newcommand{\MathEqLab}[2]{\begin{equation}\cmath{#1}\label{#2}\end{equation}}
\def\NF{\mathcal{F}} 
\def\OnticN{\mathcal{N}}
\newcommand{\ordset}[1]{\left[#1\right]} 
\newcommand{\popnumb}[1]{\left\langle#1\right\rangle} 
\def\projstd{\mathrm{P}_{\markstd}} 
\def\Q{\mathbb{Q}} 
\def\runisymb{\zeta} 
\newcommand{\runi}[1]{\runisymb_{#1}} 
\def\S{\mathbb{S}} 
\newcommand{\set}[1]{\left\{#1\right\}} 
\newcommand{\SymGN}[1]{\mathsf{S}_{#1}} 
\ifx\tr\undefined\DeclareMathOperator{\tr}{tr}\fi
\newcommand{\veco}[2]{\barket{#1}\!\brabar{#2}}	
\newcommand{\vect}[1]{\left(#1\right)} 

\title[Subsystems in Finite Quantum Mechanics]{Decomposition of a Quantum System Into\\ Subsystems in Finite Quantum Mechanics}
\author{Vladimir V. Kornyak}
\address{%
Laboratory of Information Technologies \\
Joint Institute for Nuclear Research \\
Dubna, Russia
}
\email{vkornyak@gmail.com}
\begin{document}
\begin{abstract}
Any Hilbert space with composite dimension can be factorized into a tensor product of smaller Hilbert spaces.
This allows to decompose a quantum system into subsystems.
We propose a simple tractable model for a constructive study of decompositions of quantum systems.
\end{abstract}
\vspace*{-1pt}
\maketitle
\vspace*{-6pt}
\section{Introduction}\label{intro} 
\emph{Mereology} is the study of the relations of part to whole and the relations of part to part within a whole.
\emph{Quantum mereology} studies such issues as the bipartite decomposition of a quantum system into a ``system'' and ``environment'', the relationship of a distinguished ``system'' and an ``observer'', the emergence of space and time from quantum entanglement, and other fundamental questions in quantum mechanics \cite{Carroll,Cao}.
The general scheme is as follows. 
The whole is an isolated quantum system%
\footnote{Obviously, in the exact sense, isolated systems do not exist (or they are fundamentally unobservable), with the possible exception of the Universe as a whole.}
in a given pure state undergoing a given unitary (Schr\"odinger) evolution.
Then, in a way chosen according to certain criteria, 
the system is decomposed into a tensor product of subsystems.
By reducing the ``universe'' density matrix,  we obtain mixed states for subsystems and can study the energies and entanglement measures associated with subsystems, and their time evolution.
For the corresponding computations, involving rather tedious combinatorics, we develop a model based on a finite formulation of quantum mechanics.
\section{Factorization of a Hilbert Space}\label{factor} 
\paragraph{Tensor product of Hilbert spaces}
The (\emph{global}) Hilbert space \Math{\Hspace} of a \Math{K}-component quantum system is the tensor product of the (\emph{local}) Hilbert spaces \Math{\Hspace_k} of the components:
\MathEqLab{\Hspace=\KroneckerProduct_{k=1}^K\Hspace_k\,.}{HspaceN}
If \Math{\dim\Hspace=\OnticN} and \Math{\dim\Hspace_k=d_k}, then \Math{\OnticN=\prod_{k=1}^K{}d_k}.
For any \Math{d}-dimensional Hilbert space, we denote the \Math{i}th orthonormal basis element by \Math{\barket{i}}, that is, 
\Math{\barket{0}=\vect{1,0,\ldots}^{\top}},
 \Math{\barket{1}=\vect{0,1,0,\ldots}^{\top}},\Math{\ldots,\barket{d-1}=\vect{0,0,\ldots,1}^{\top}}.
Tensor monomials of local basis elements form an orthonormal basis in the global Hilbert space:
\MathEqLab{\barket{i}=\barket{i_1}\KroneckerProduct\cdots\KroneckerProduct\barket{i_k}\KroneckerProduct\cdots\KroneckerProduct\barket{i_K},}{1to1corresp}
where \Math{\barket{i}\in\Hspace}, \Math{\barket{i_k}\in\Hspace_k} and
\MathEqLab{i=i_1\prod_{m=2}^Kd_m+\ldots+i_k\prod_{m=k+1}^Kd_m+\ldots+i_K.}{1to1numeric} 
\paragraph{Tensor decomposition of a Hilbert space}
We can {reverse the procedure}, since \eqref{1to1corresp} is a one-to-one correspondence --- the sequence \Math{i_1,\ldots,i_K} is uniquely recovered from \Math{i} using  formula \eqref{1to1numeric}.
\par
Starting with an orthonormal basis in an \Math{\OnticN}-dimensional Hilbert space \Math{\Hspace} and a decomposition \Math{\OnticN=d_1\cdots{}d_K}, 
we can construct a particular isomorphism between \Math{\Hspace} and the tensor product of local spaces of the corresponding dimensions.
\par
When constructing an isomorphism, we must take into account the freedom in the choice of bases in Hilbert spaces.
Any two orthonormal bases are related by a unitary transformation.
Using the properties of the tensor product, we can write 
\Ali{&U\barket{\psi}=\,U_1\barket{\psi_1}\KroneckerProduct\cdots\KroneckerProduct{}U_K\barket{\psi_K}
=\vect{U_1\KroneckerProduct\cdots\KroneckerProduct{}U_K}\vect{\barket{\psi_1}\KroneckerProduct\cdots\KroneckerProduct\barket{\psi_K}}\\
&\implies\vect{U_1\KroneckerProduct\cdots\KroneckerProduct{}U_K}^{-1}U\barket{\psi}=\barket{\psi_1}\KroneckerProduct\cdots\KroneckerProduct\barket{\psi_K},
}
\!\!where \Math{\barket{\psi}\in\Hspace,\, \barket{\psi_k}\in\Hspace_k;\,U,\,U_k} are unitary transformations in the corresponding spaces.
We see that local transformations can be absorbed by the global transformation \Math{U},%
~so in general we have
\MathEq{U\barket{\psi}=\barket{\psi_1}\KroneckerProduct\cdots\KroneckerProduct\barket{\psi_K}.} 
Thus, to specify a factorization of a Hilbert space \Math{\Hspace} we need a \emph{decomposition} of \Math{\dim\Hspace} and a \emph{unitary transformation} that fixes a basis in  \Math{\Hspace}.
\section{Finite Version of Quantum Mechanics}\label{FQM} 
We use a version of quantum theory \cite{Kornyak17ISQS,Kornyak17MMCP,Kornyak11MMCP} in which the groups of unitary evolutions are replaced by linear representations of finite groups, and the field of complex numbers is replaced by its dense constructive subfields that naturally arise from the natural numbers and roots of unity.
\paragraph{Permutation Hilbert space}
The fact that \emph{any} linear (hence \emph{unitary}) representation of a finite group is a subrepresentation of some permutation representation implies that the formalism of quantum mechanics can be completely%
\footnote{Modulo empirically insignificant elements of traditional formalism such as infinities of various kinds.} reproduced based on permutations of some set 
\MathEqLab{\Omega=\set{e_1,\ldots,e_\mathcal{N}}\cong\set{1,\ldots,\mathcal{N}}}{omega}
of primary (\emph{``ontic''}) objects on which a permutation group \Math{G\leq\SymGN{\OnticN}} acts.
\par
The Hilbert space on the set \Math{\Omega}, necessary for calculating quantum probabilities, can be most economically constructed on the basis of two primitive concepts: (a) \emph{natural numbers}  \Math{\N=\set{0,1,\ldots}}, abstraction of \emph{counting},   and (b) \emph{roots of unity}, abstraction of \emph{periodicity}.
\par
To construct a field \Math{\NF} sufficient for all calculations in the quantum formalism, in particular, for splitting any representation of any subgroup of  \Math{G} into irreducible components, we can proceed as follows.
We extend the semiring \Math{\N} to the ring  \Math{\N\ordset{\runi{\ell}}},
where \Math{\runi{\ell}} is the \Math{\ell}th primitive root of unity, and  \Math{\ell} is the least common multiple of the orders (periods) of the elements of  \Math{G}.
The \emph{algebraic integer} \Math{\runi{\ell}} can be written in complex form as \Math{\displaystyle\runi{\ell}=\e^{{2\pi}\imu/{\ell}}}.
Finally, constructing the \emph{field of fractions} of the ring \Math{\N\ordset{\runi{\ell}}}, we arrive at the \emph{cyclotomic extension} of the rationals \MathEq{\NF=\Q\farg{\e^{{2\pi}\imu/{\ell}}}.}
For \Math{\ell>2}, the field \Math{\NF}, being a dense subfield of \Math{\C}, is physically (empirically) indistinguishable from the field of complex numbers. 
\par
Treating the set \Math{\Omega} as a basis, we obtain an \Math{\OnticN}-dimensional Hilbert space \Math{\Hspace_\OnticN} over the field \Math{\NF}.
The action of \Math{G} on \Math{\Omega} determines the \emph{permutation representation} \Math{\mathcal{P}} in  \Math{\Hspace_\OnticN} by the matrices
\MathEq{\mathcal{P}\farg{g}_{i,j}=\delta_{ig,j},} 
where \Math{ig} denotes the (right) action of \Math{g\in{}G} on \Math{i\in\Omega}.
\paragraph{Decomposition of permutation representation}
The permutation representation of any group \Math{G} has the \emph{trivial} one-dimensional subrepresentation in the space spanned by the all-ones vector
\MathEq{\barket{\omega}=(\underbrace{1,1,\ldots,1}_{\OnticN})^{\top}.}
The complement to the trivial subrepresentation is called the \emph{standard representation}.
The operator of projection onto the \Math{\vect{\OnticN-1}}-dimensional \emph{standard space} \Math{\Hspace_{\,\markstd}} has the form
\MathEq{\projstd=\idmat_{\OnticN} - \frac{\veco{\omega}{\omega}}{\OnticN}\,.}
Quantum mechanical behavior (interference, etc.) manifests itself precisely in the standard representation. 
Banks made a profound observation \cite{Banks} that the projection of classical permutation evolutions in the whole \Math{\Hspace_\OnticN} leads to truly quantum evolutions in the subspace \Math{\Hspace_{\,\markstd}} and demonstrated that the choice \Math{G=\SymGN{\OnticN}}, where \Math{\OnticN} is the number of fundamental (Planck) elements,%
\footnote{The number  \Math{\OnticN} is estimated as \Math{
\sim\Exp\farg{\Exp\farg{20}}} and \Math{
\sim\Exp\farg{\Exp\farg{123}}} for 1 cm$^3$ of matter and for the entire Universe, respectively.}
makes it possible to reproduce finite dimensional approximations to all known models of theoretical physics.
\paragraph{Quantum states as projections of natural vectors}
\Math{\SymGN{\OnticN}} is a \emph{rational-representation group}, i.e., its every irreducible representation (the standard representation is one of them) is realizable over \Math{\Q}.
This means that to describe evolutions in \Math{\Hspace_{\,\markstd}}, it is sufficient to consider only vectors with rational components.%
\footnote{Complex numbers (nontrivial elements of cyclotomic extensions) can be required only when a representation of a proper subgroup of \Math{\SymGN{\OnticN}} must be split into irreducible components.}
Taking into account that a quantum state is a ray in vector space, it is easy to show that any quantum state in \Math{\Hspace_{\,\markstd}} can be obtained by projection of vectors from \Math{\Hspace_\OnticN} with natural components. 
These \emph{natural} vectors are integer points of the nonnegative orthant:
\Math{\barket{n}=\vect{n_1,\ldots,n_\OnticN}^\top\in\N^{\OnticN}\subset\Hspace_\OnticN.}
To build constructive models, one need to select a finite subset in \Math{\N^{\OnticN}}.
Natural vectors whose coordinates belong to the set \Math{\set{0,1,\ldots,m}, m\geq2,} will be called \Math{m}\emph{th order vectors}.
Quantum states described by collinear vectors are equivalent.
To eliminate this redundancy, we can pre-project the natural vectors onto  \Math{\S^{^{\OnticN-1}}_+}\!\!, the part of the unit sphere in the nonnegative orthant.
The total number of quantum states defined by vectors of the \Math{m}th order is not less than
\MathEqLab{2^{\OnticN}-2\,.}{Rm}
The area of \Math{\S^{^{\OnticN-1}}_+} is equal to
\MathEqLab{\frac{\OnticN\pi^{\OnticN/2}}{2^{\OnticN}\Gamma\farg{\OnticN/2+1}}\approx\sqrt{\frac{\OnticN}{\pi}}\vect{\frac{\e\pi}{2\OnticN}}^{\OnticN/2}\!.}{area}
A rough comparison of the exponentially growing --- as can be seen from \eqref{Rm} --- number of quantum states with the rapidly decreasing area \eqref{area} shows that for large \Math{\OnticN}, the \Math{m}th order vectors represent a significant part of quantum states.
\paragraph{Ontic vectors}
We call \Math{2}nd order vectors \emph{ontic vectors}.
These vectors are attractive for both ontological and computational reasons.
The ontic vector \Math{\barket{q}} can be written as a bit string of length \Math{\OnticN}.
Interpreting this string as a \emph{characteristic function}, one can identify the ontic vector with the corresponding nontrivial subset of the set of ontic elements \eqref{omega}:
 \Math{q\subset\Omega}. 
The complete set of ontic vectors is \MathEq{Q=2^\Omega\setminus\set{\emptyset,\Omega},~~ \cabs{Q}=2^\OnticN-2\,.}
\indent{}The inner product of ontic vectors \Math{\barket{q}} and \Math{\barket{r}} in the space \Math{\Hspace_\OnticN} has the form \Math{\inner{q}{r}=\popnumb{q\bitand{}r}}, 
where \Math{\bitand} is the bitwise AND for bit strings, and \Math{\popnumb{\,\cdot\,}} is the number of ones in a bit string (\emph{population number} or \emph{Hamming weight}).
For the inner product of normalized projections of  \Math{\barket{q}} and \Math{\barket{r}} onto \Math{\Hspace_{\,\markstd}}, we have
\MathEq{S\farg{q,r}\equiv\frac{\innerform{q}{\projstd}{r}}{\sqrt{\innerform{q}{\projstd}{q}\innerform{r}{\projstd}{r}}}
=\frac{\OnticN\!\popnumb{q\,\bitand\,r}-\!\popnumb{q}\popnumb{r}}{\sqrt{\popnumb{q}\popnumb{\bitnot\!q}\popnumb{r}\popnumb{\bitnot\!r}}},}
where \Math{\bitnot} denotes bitwise inversion.
The obvious identities
\Math{\popnumb{\bitnot\!a}=\OnticN-\popnumb{a}} and \Math{\popnumb{a\,\bitand\,b}+\popnumb{a\,\bitand\bitnot\!b}=\popnumb{a}}
imply the folowing symmetry with respect to transpositions of subsets of the ontic set and their complements
\MathEq{S\farg{q,r}=-S\farg{\bitnot\!q,r}=-S\farg{q,\bitnot\!r}=S\farg{\bitnot\!q,\bitnot\!r}\,.}
\section{Decomposition of a Quantum System}\label{Decomposition} 
Since any mixed state of a quantum system can be obtained from a pure state in a larger Hilbert space by taking a partial trace, it is natural to assume that at a fundamental level the state of an isolated system must be pure.%
\footnote{This belief is expressed by the metaphor ``the Church of the Larger Hilbert Space'' (J.A. Smolin).}
\par
The original permutation basis in the space \Math{\Hspace_{\OnticN}}, i.e., the set \Math{\Omega}, will be referred as the \emph{ontic basis}.
In this basis, the pure density matrix in \Math{\Hspace_{\,\markstd}} associated with the ontic state \Math{\barket{q}\in\Hspace_{\OnticN}} has the form
\MathEqLab{\denmat_q^o=\frac{\projstd\veco{q}{q}\projstd}{\innerform{q}{\projstd}{q}}=\frac{1}{\OnticN}
\frac{\vect{\barket{q}-\alpha\barket{\omega}}
\vect{\brabar{q}-\alpha\brabar{\omega}}}
{\alpha\vect{1-\alpha}}\,,}{Rho}
where \Math{\displaystyle\alpha=\frac{\popnumb{q}}{\OnticN}} is the \emph{population density}.
There is an obvious duality: the expression for the density matrix \Math{\displaystyle\denmat_{\bitnot{}q}^o} is obtained from \eqref{Rho} by replacements 
\Math{q\rightarrow\,\,\bitnot\hspace{-4pt}q} and \Math{\alpha\rightarrow1-\alpha}.
\paragraph{Energy basis}
In continuous QM, the evolution of an isolated system is described by the one-parameter unitary group \Math{U_t=\e^{-\imu{}Ht}} generated by the Hamiltonian \Math{H} whose eigenvalues are called \emph{energy eigenvalues}.
In finite QM, the evolution is described by a unitary representation of a cyclic group \Math{U\farg{g}^t} generated by an element \Math{g\in{}G}, where \Math{t} is an integer parameter.
In our case, the generator of evolution is some matrix \Math{\mathcal{P}\farg{g}} from the permutation representation of  \Math{G} in \Math{\Hspace_\OnticN}.
We call the \emph{energy basis} an orthonormal basis in which the matrix \Math{\mathcal{P}\farg{g}} is diagonal.
\par
Any permutation can be written as a product of disjoint cycles.
It is easy to show that the total number of cycles of length \Math{\lcycle} in the whole group \Math{\SymGN{\OnticN}} is \Math{\displaystyle\frac{\OnticN!}{\lcycle}},
and, therefore, the expected number of \Math{\lcycle}-cycles in a single permutation is \Math{\displaystyle\frac{1}{\lcycle}}.
That is, high-frequency (high-energy) evolutions are more common.
\par
The \Math{\lcycle}-cycle matrix has the form
\MathEq{C_{\lcycle}=
\bmat
0&1&0&\cdots&0\\
0&0&1&\cdots&0\\
\vdots&\vdots&\vdots&\vdots&\vdots\\
1&0&0&\cdots&0
\emat\,.} 
The diagonal form of this matrix is
	\MathEq{
	F_{\lcycle}C_{\lcycle}F^{-1}_{\lcycle}=
\bmat
1&0&0&\cdots&0\\
0&\runi{\lcycle}&0&\cdots&0\\
0&0&\runi{\lcycle}^2&\cdots&0\\
\vdots&\vdots&\vdots&\vdots&\vdots\\
0&0&0&\cdots&\runi{\lcycle}^{\lcycle-1}
\emat\,,}
where \Math{\displaystyle\runi{\lcycle}=\e^{{2\pi}\imu/{\lcycle}}} is the \Math{\lcycle}th primitive root of unity, and 
\MathEq{F_{\lcycle}=\frac{1}{\sqrt{\lcycle}}
\bmat
1&1&1&\cdots&1\\
1&\runi{\lcycle}^{-1}&\runi{\lcycle}^{-2}&\cdots&\runi{\lcycle}^{-\vect{\lcycle-1}}\\
1&\runi{\lcycle}^{-2}&\runi{\lcycle}^{-4}&\cdots&\runi{\lcycle}^{-2\vect{\lcycle-1}}\\
\vdots&\vdots&\vdots&\cdots&\vdots\\
1&\runi{\lcycle}^{-\vect{\lcycle-1}}&\runi{\lcycle}^{-2\vect{\lcycle-1}}&\cdots&\runi{\lcycle}^{-\vect{\lcycle-1}\vect{\lcycle-1}}
\emat}
is the \emph{Fourier transform} matrix. 
\Math{F_{\lcycle}} is both unitary and symmetric, therefore \Math{\displaystyle{}F^{-1}_{\lcycle}=F_{\lcycle}^*.} 
In general, the matrix of the permutation representation of an element \Math{g\in{}\SymGN{\OnticN}} is the direct sum of cyclic matrices
\Math{\displaystyle{}\mathcal{P}\farg{g}=\bigoplus\limits_{m=1}^MC_{\lcycle_m}}, and the corresponding diagonalizing matrix is \Math{\displaystyle{}F=\bigoplus\limits_{m=1}^MF_{\lcycle_m}},
which is the transition matrix from the \emph{ontic basis} to the \emph{energy basis}.
The density matrix in the energy basis can be calculated from \eqref{Rho} by the formula
\MathEq{\denmat_q^e=F\denmat_q^oF^{-1}.}
\paragraph{Decomposition of a quantum state}
If the basis of a space \Math{\Hspace} is fixed, then the procedure for the decomposition of a quantum system in the state \Math{\denmat} is reduced to the following:
\begin{enumerate}
	\item The factorization of \Math{\dim\Hspace} is chosen: \Math{\dim\Hspace=d_1\cdot{}\ldots\cdot{}d_K}.
	\item Subsystems are identified with subsets \Math{A} of the set of points \MathEqLab{X=\set{1,\ldots,K}.}{setX}
	~\\[-20pt]
	\item 
The basis elements of \Math{\Hspace} are identified with the tensor monomials from the tensor product 
\eqref{HspaceN} in accordance with \eqref{1to1corresp} and using formula \eqref{1to1numeric}.	
	\item 
The density matrix of the subsystem \Math{A} is calculated by taking the trace over the complement:	\Math{\denmat_A=\tr_{X\setminus{}A}\denmat}.
\end{enumerate}
Having density matrices of subsystems, it is possible to calculate the physical characteristics describing subsystems, and interactions and quantum correlations between them:
energies, entropies, mutual information and other entanglement measures.
\section{Entanglement Measures}\label{Measures}
Quantitatively, quantum correlations are described by \emph{measures of entanglement}, which are based on the concept of entropy.
The most commonly used in physics is the \emph{von Neumann entropy}
\Math{S_1\farg{\denmat}=-\tr\farg{\denmat\log\denmat}.} 
Also often used are entropies from the \emph{R\'enyi family}
\MathEq{S_\alpha\farg{\denmat}=\frac{1}{1-\alpha}\log\tr\farg{\denmat^{\alpha}},~~\alpha\geq0,~\alpha\neq1.} 
The common feature of the von Neumann and R\'enyi entropies is their additivity on combinations of independent probability distributions determined by the eigenvalues of the density matrices.
The von Neumann entropy is preferred because it additionally satisfies a stronger requirement, the chain rule for conditional entropies.
\par
In our calculations, we use the 2nd R\'enyi entropy (also called the \emph{collision entropy}) \Math{S_2\farg{\denmat}=-\log\tr\farg{\denmat^2}} for the following reasons:
\begin{itemize}
	\item
It is easy to calculate: \Math{\tr\farg{\denmat^2}=\sum\limits_{i=1}^n\denmat_{ii}^2+2\sum\limits_{i=1}^{n-1}\sum\limits_{j=i+1}^{n}\cabs{\denmat_{ij}}^2.}
	\item
The value \Math{\tr\farg{\denmat^2}} coincides with the Born probability: ``the system observes itself.'' 
	\item 
The value \Math{\tr\farg{\denmat^2}} is the square of the \emph{Frobenius} (\emph{Hilbert-Schmidt})  \emph{norm} of the density matrix.
\end{itemize}
\section{\emph{\emph{Illustrative Calculation}}}\label{Illustrative} 
\begin{center}
\includegraphics[width=0.95\textwidth,height=0.3998\textheight]{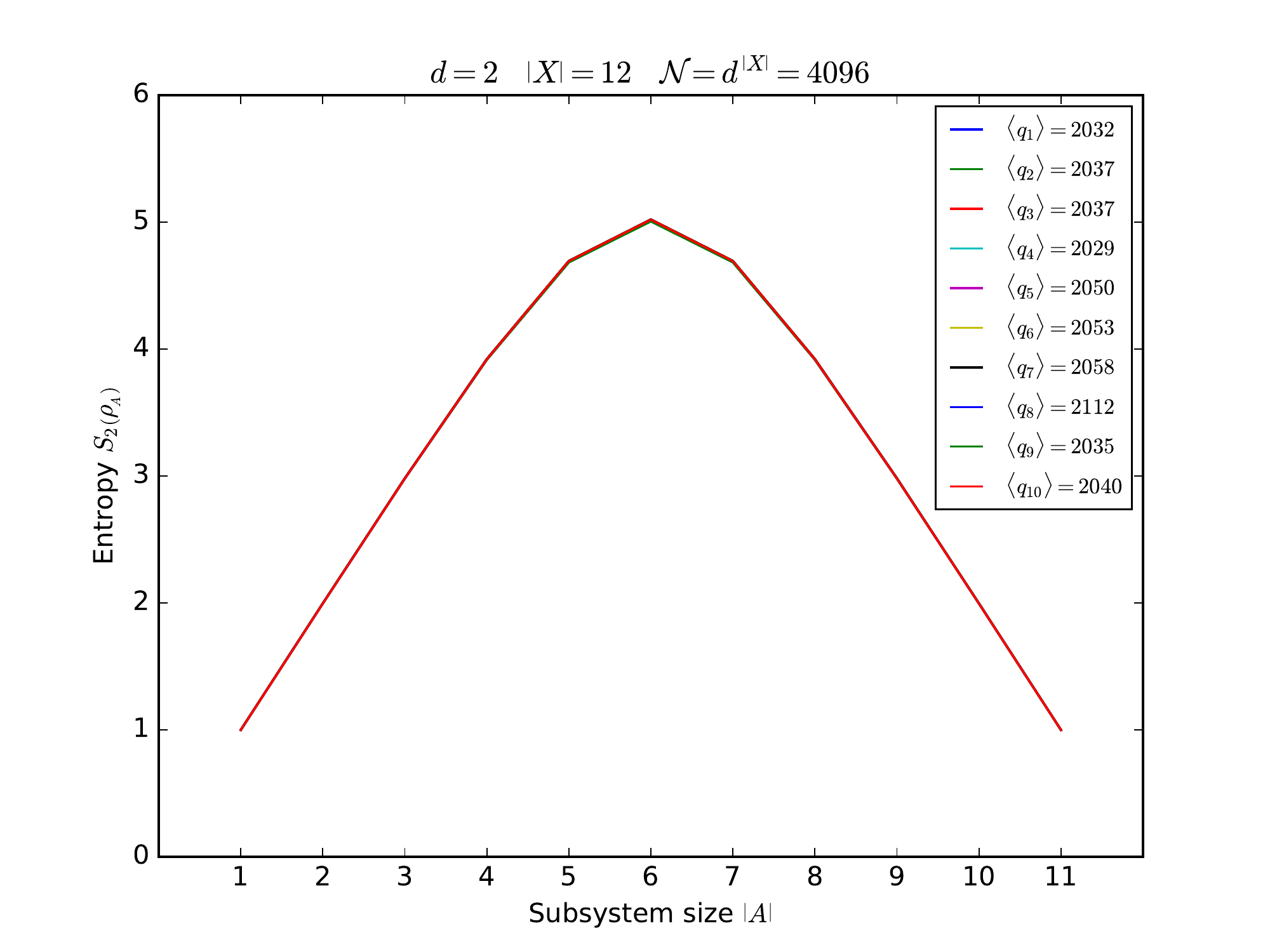}
\end{center}
The figure presents the values of entropy \Math{S_2\farg{\denmat_A}} computed in the ontic basis for the decomposition \Math{\OnticN=2^{12}}.
Data for subsystems of all possible sizes, computed for ten randomly generated ontic vectors, demonstrate the following features:
\begin{itemize}
	\item
Weak dependence on the quantum state: visually, all graphs are almost identical.
Note that this behavior arises for a sufficiently large number of the decomposition components \eqref{setX}. 
In this case, \Math{\cabs{X} = 12}.
	\item 
Symmetry \Math{S_2\farg{\denmat_A}=S_2\farg{\denmat_{X\setminus{}A}}}	is a manifestation of the \emph{Schmidt} bipartite \emph{decomposition} of a pure state:
both matrices \Math{\denmat_A} and \Math{\denmat_{X\setminus{}A}} have identical sets of nonzero eigenvalues.
	\item 
For \Math{\cabs{A}} noticeably smaller than \Math{\cabs{X}/2}, the reduced state is close to the \emph{maximally mixed} state:	\Math{S_2\farg{\denmat_{_A}}\approx\cabs{A}\log{}d}.
In our example, \Math{d=2}.
\end{itemize}

\end{document}